\documentclass[aps,a4paper,groupedaddress,floatfix,showpacs,longbibliography]{revtex4-1}
\usepackage{graphicx,amsmath,amssymb}
\usepackage{subeqnarray}

\newcommand{\rin}{r_1}
\newcommand{\rout}{r_2}

\newcommand{\bu}{\mathbf{u}}
\newcommand{\bU}{\mathbf{U}}

\newcommand{\be}{\mathbf{e}}
\newcommand{\pd}{\partial}
\newcommand{\divv}{\nabla\cdot}

\newcommand{\curl}{\nabla\times}
\newcommand{\lap}{\nabla^2}
\newcommand{\gap}{d}

\newcommand{\wave}{L}
\newcommand{\MP}{M_0}

\begin{document}
\title{The Self-Sustaining Process in Taylor-Couette Flow}
\author{Tommy Dessup}
%\affiliation{MSC, CNRS (UMR XXXX), Univ. Paris 7, 75013 Paris, France}
\affiliation{Physique et M\'ecanique des Milieux 
H\'et\'erog\`enes (PMMH),
CNRS, ESPCI Paris, PSL Research University, Sorbonne Universit\'e, 
Univ. Paris Diderot, 75005 France}
\author{Laurette S. Tuckerman}
\affiliation{Physique et M\'ecanique des Milieux 
H\'et\'erog\`enes (PMMH),
CNRS, ESPCI Paris, PSL Research University, Sorbonne Universit\'e, 
Univ. Paris Diderot, 75005 France}
\author{Jos\'e Eduardo Wesfreid}
\affiliation{Physique et M\'ecanique des Milieux 
H\'et\'erog\`enes (PMMH),
CNRS, ESPCI Paris, PSL Research University, Sorbonne Universit\'e, 
Univ. Paris Diderot, 75005 France}
\author{Dwight Barkley}
\affiliation{Mathematics Institute, University of Warwick, Coventry CV4 7AL,
  United Kingdom}
\author{Ashley P. Willis}
\affiliation{School of Mathematics and Statistics, University of 
Sheffield, Sheffield, S3 7RH United Kingdom}
\date{\today}

\begin{abstract}
  The transition from Tayor vortex flow to wavy-vortex flow is revisited.
  The Self-Sustaining Process (SSP) of Waleffe [Phys. Fluids {\bf 9}, 883--900
    (1997)]
  proposes that a key ingredient in
  transition to turbulence in wall-bounded shear flows is a three-step process
  involving rolls advecting streamwise velocity, leading to streaks which
  become unstable to a wavy perturbation whose nonlinear interaction with
  itself feeds the rolls. We investigate this process in Taylor-Couette flow.
  The instability of Taylor-vortex flow to wavy-vortex flow, a process which
  is the inspiration for the second phase of the SSP, is shown to be caused 
  by the streaks, with the rolls playing a negligible role,
  as predicted by Jones [J. Fluid Mech. {\bf 157}, 135--162 (1985)]
  and demonstrated by Martinand et al.~[Phys. Fluids {\bf 26}, 094102 (2014)].
  In the third phase of the SSP, the nonlinear interaction of the
  waves with themselves reinforces the rolls. We show this both 
  quantitatively and qualitatively, identifying physical regions in
  which this reinforcement is strongest, and also demonstrate that
  this nonlinear interaction depletes the streaks.
\end{abstract}


\maketitle

\parindent .5cm

\section{Introduction}
\label{sec:Introduction}

The first successful linear stability analysis for a viscous fluid was carried
out in 1923 by G.I. Taylor \cite{taylor1923stability} for the flow between two
concentric differentially rotating cylinders.  What then became known as
Taylor-Couette flow has played a central role in hydrodynamic stability theory
ever since.  In the standard configuration of a stationary outer cylinder, as
the inner cylinder rotation rate is increased, laminar flow is succeeded by
axisymmetric Taylor vortices via the centrifugal instability first explained
by Rayleigh \cite{rayleigh1916}.  The Taylor vortices subsequently develop
azimuthal waves, seen in experiments by researchers such as Coles
\cite{coles1965transition}, Swinney and co-workers
\cite{gorman1982spatial,king1984wave,andereck1986flow}
and others \cite{hegseth1996bifurcations,wereley1998spatio}. Wavy-vortex
flow was studied computationally when this became possible in the 1980s by
authors such as Jones \cite{jones1981nonlinear,jones1985transition}, Marcus
\cite{marcus1984a,marcus1984b} and others \cite{edwards1991onset,antonijoan2002stable}.

Taylor-Couette flow has also been studied as a way of approaching plane
Couette flow, which undergoes transition to three-dimensional turbulence 
despite being linearly stable at all Reynolds numbers.  The azimuthal, radial,
and axial directions of Taylor-Couette flow play the role of the streamwise,
cross-channel, and spanwise direction, respectively. As the ratio between the
cylinder radii approaches one, the correspondence between the two flows
becomes exact.  The possibility of approaching plane Couette flow via
Taylor-Couette flow has been used by many authors for many different purposes.
Nagata \cite{nagata1990three} used homotopy to calculate otherwise
inaccessible unstable steady states of plane Couette flow. Hristova et al.
\cite{hristova2002transient} and Meseguer et al. \cite{meseguer2002energy}
compared transient growth rates between the two flows.  Prigent et
al.~\cite{prigent2002large} extended the observation of coexisting turbulent
and laminar regions seen in Taylor-Couette by Coles \cite{coles1965transition}
to plane Couette flow.  Faisst and Eckhardt \cite{faisst2000transition} used
Taylor-Couette flow to approach the turbulent lifetimes and intermittency of
plane Couette flow.  A very narrow gap Taylor-Couette geometry was used as a
proxy for plane Couette flow by Shi et al. \cite{shi2013scale} to calculate
the statistical threshold of sustained turbulence and by Lemoult et
al. \cite{lemoult2016directed} to establish that this transition was
manifested as a directed percolation phase transition.

Here, we take the analogy in the opposite direction: extending an idea
developed for plane Couette flow to Taylor-Couette flow.  Waleffe
\cite{waleffe1995hydrodynamic,hamilton1995regeneration,waleffe1997self}
has proposed a now widely-accepted three-part mechanism, by which
streamwise rolls (damped in the plane Couette case), cause streamwise
streaks (by simple advection of the streamwise velocity contours),
which become wavy (through instability), acquiring streamwise
dependence.  The nonlinear self-interaction of the wavy streaks drives
the streamwise rolls, thus closing the cycle.  The mechanism is
similar to that proposed by Hall and co-workers
\cite{hall1991strongly,hall2010streamwise,blackburn2013lower} and by
Beaume and co-workers
\cite{beaume2014exact,beaume2015reduced,beaume2016modulated}.
Experimental evidence for the SSP in plane boundary layer and channel
flow has been reported by Wesfreid and colleagues in
\cite{duriez2009self,klotz2017experiments}.  These experiments show a
strong correlation between the growth of rolls and the presence of
waves: both phenomena occur above the same Reynolds-number threshold.

Although the SSP was influenced by these phenomena in Taylor-Couette flow, it
has not actually been applied to Taylor-Couette flow itself.  The main
purpose of this paper is to see how the Self-Sustaining Process plays out in
Taylor-Couette flow, where the analogous structures, i.e. axisymmetric and
wavy Taylor vortices, are actually stable equilibrium states.

\section{Equations, Methods and Parameters}

The equations governing Taylor-Couette flow and the methods for computing it
are sufficiently well known as to warrant only a very brief exposition.  The
inner and outer cylinders have radii and angular velocities $R_j$ and
$\Omega_j$.  From these, along with the kinematic viscosity $\nu$, can be
constructed the length scale $\gap \equiv R_2-R_1$, the time scale
$\gap^2/\nu$, the two Reynolds numbers $Re_j\equiv R_j\Omega_j\gap/\nu$, and
the radius ratio $\eta\equiv R_1/R_2$.  The non-dimensionalized governing
equations and boundary conditions are then
\begin{subequations}
\begin{align}
\pd_t \bU &= \bU\times\curl\bU  -\nabla P + \lap\bU \\
  \divv \bU &= 0 \\
  \bU &= Re_j \be_\theta\mbox{~at~} r=r_j \equiv R_j/d,\,\, j=1,2
\end{align}
\label{eq:NS}\end{subequations}
We will restrict our consideration to the classic inner-cylinder-rotation case
with $\Omega_2=0$ so that $Re_2=0$ and hence we use $Re$ to denote the
inner Reynolds number $Re_1$.  
Nonlinear Taylor-vortex and wavy-vortex flows, denoted by TVF and WVF or
$\bU_{\rm TVF}$ and $\bU_{\rm WVF}$, are calculated by solving the evolution
equations \eqref{eq:NS} numerically.
For linear stability analysis, the nonlinear code has been adapted to solve
the linearized equations
\begin{subequations}
\begin{align}
\pd_t \bu &= \bU\times\curl\bu + \bu\times\curl \bU -\nabla p + \lap\bu \\
  \divv \bu &= 0 \\
  \bu &= 0 \quad\mbox{~at~} r=r_j,\,\, j=1,2
\end{align}
\label{eq:linstab}\end{subequations}
where $\bU$ is the flow whose stability is sought.
Temporal integration of \eqref{eq:linstab} effectively carries out the power
method, converging to the eigenvector whose eigenvalue has largest real part.
Most commonly, we take $\bU$ to be Taylor-vortex flow, $\bU_{\rm TVF}$, and 
the power method returns the wavy vortex eigenvector $\bu_{\rm wvf}$ and
corresponding eigenvalue. 

The code we use represents functions on a spatial Chebyshev
grid in the radial direction $r$ and on equally spaced points in the azimuthal
$\theta$ and axial $z$ directions, with spatial derivatives taken via finite
differences in $r$ and by differentiation of Fourier series in $\theta$, $z$.
Multiplications are carried out in the grid space representation
by Fourier transforming in $\theta,z$.
Taylor-vortex flow is calculated in an axisymmetric domain with $N_r=33$
radial points and $N_z=16$ points over the axial domain $[0,\wave_z]$ or,
equivalently, multiples of the wavenumber $2\pi/\wave_z$.  Computations of
wavy-vortex flow eigenvectors use a single azimuthal mode $\MP$.  Nonlinear
wavy-vortex flow is calculated using $N_\theta=16$ points in the azimuthal
sector $[0,2\pi/\MP]$ or, equivalently, multiples of the wavenumber $\MP$.

\begin{figure}[t]
\includegraphics[width=0.85\columnwidth]{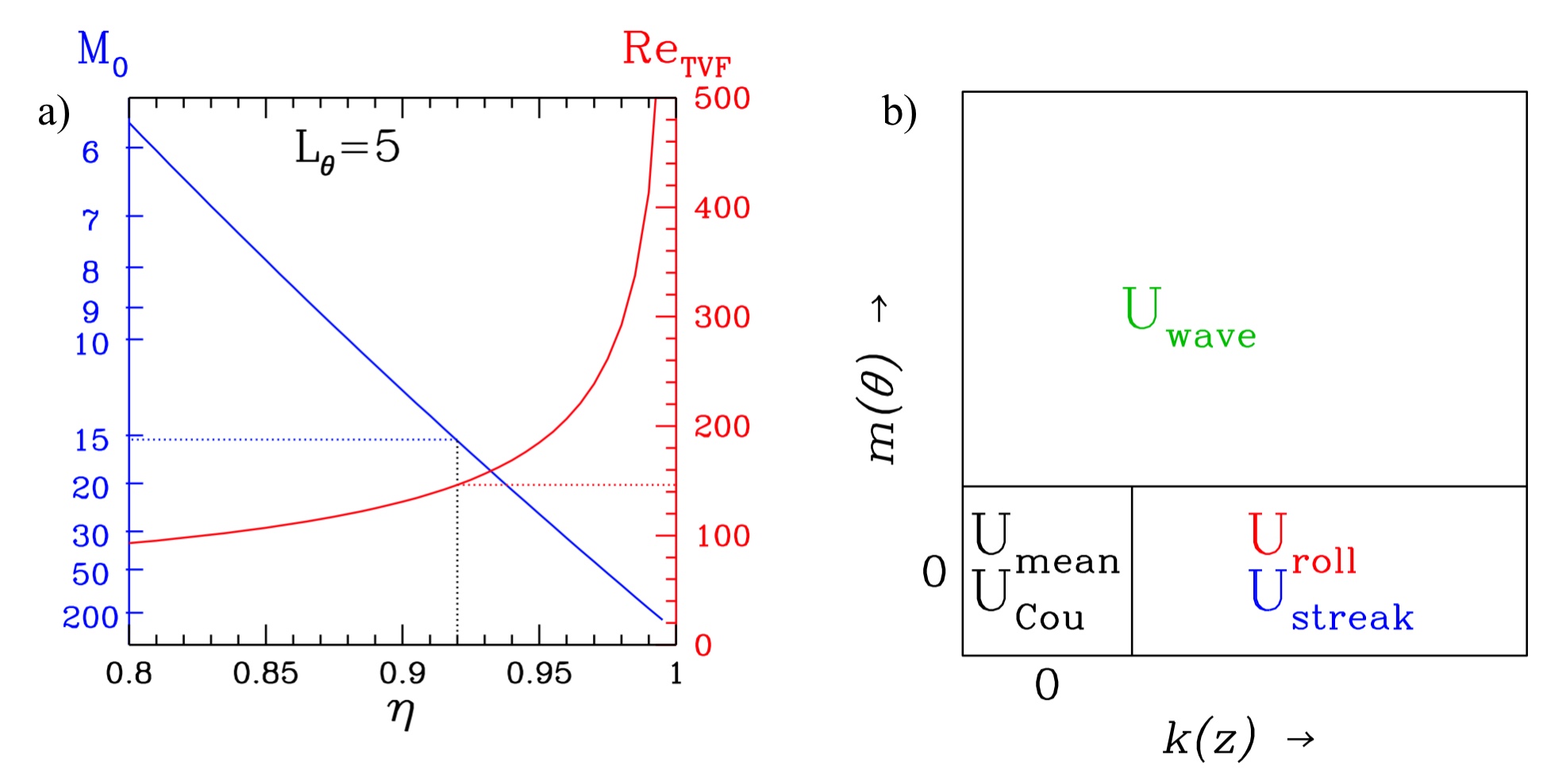}
  \caption{(a) Dependence of critical Reynolds number $Re^{\rm TVF}$ on the
    radius ratio $\eta$ for transition to Taylor-vortex flow (right
    scale). Also shown is the relationship between azimuthal wavenumber $\MP$
    and $\eta$ for circumferential wavelength $L_\theta=5$ (left scale).  Our
    chosen parameter values are $\eta=0.92$ and $\MP=15$.  (b) Schematic
    decomposition of flow into $\bU_{\rm Cou}$, $\bU_{\rm mean}$, $\bU_{\rm roll}$,
    $\bU_{\rm streak}$, $\bU_{\rm wave}$ according to axial and azimuthal Fourier
    modes $k$ and $m$.}
  \label{fig:detpar}
\end{figure}

One difficulty is deciding which of the many TVFs or WVFs to study. Each TVF
is characterized by an axial wavenumber, and each WVF has an axial and an
azimuthal wavenumber.  States with different wavenumbers can be simultaneously
stable, as emphasized by Coles~\cite{coles1965transition} and by many
subsequent researchers \cite{gorman1982spatial,king1984wave,hegseth1996bifurcations}.
Jones \cite{jones1985transition} and Antonijoan \& Sanchez
\cite{antonijoan2002stable} have shown the complexity of the bifurcations and
ranges of existence of wavy-vortex states with different azimuthal wavenumbers
as the radius ratio and the axial wavelength are varied.  We select the radius
ratio to be $\eta=0.92$, corresponding to $r_1=11.5$ and $r_2=12.5$.
To make connection with the SSP in plane Couette flow, we take the axial
wavelength to be $\wave_z=2$, corresponding to a spanwise wavelength of 4
half-gaps, near the length considered by Waleffe
\cite{waleffe1995hydrodynamic,hamilton1995regeneration,waleffe1997self}.
(Note that the length scale in
the Taylor-Couette problem is the full gap.)
We use the term circumferential wavelength to denote a length at the midgap $r=\bar{r}$, in contrast
with an azimuthal wavelength, which is expressed in radians and necessarily a
fraction of $2\pi$.
To approximate the streamwise wavelength of 10 half-gaps studied by Waleffe,
we first express the circumferential wavelength $L_\theta$ of a wavy-vortex state with
azimuthal wavenumber $\MP$ in units of the gap
\begin{equation}
  \wave_\theta = \frac{2\pi \bar{r}}{\MP}
  =\frac{2\pi}{\MP}\;\frac{(\rout+\rin)/2}{\rout-\rin}
  =\frac{\pi}{\MP}\;\frac{1+\frac{\rin}{\rout}}{1-\frac{\rin}{\rout}}
  =\frac{\pi}{\MP}\;\frac{1+\eta}{1-\eta}
\end{equation}
Setting $\eta=0.92$ and $\wave_\theta=5$, corresponding to 10 in half-gaps,
leads to
\begin{equation}
   \MP=\frac{\pi}{\wave_\theta}\;\frac{1+\eta}{1-\eta}
 =\frac{\pi}{5}\;\frac{1.92}{0.08}\approx 15
\end{equation}
The critical Reynolds number for onset of Taylor-vortex flow in which only the
inner cylinder rotates is approximately
\begin{equation}
  Re_{\rm TVF}\approx\sqrt{\frac{1708}{\eta(1-\eta)}}\frac{1+\eta}{2}
\end{equation}  
which diverges as the narrow-gap (or plane Couette) limit $\eta\rightarrow 1$
is approached \cite{faisst2000transition}.
The dependence of the critical Reynolds on $\eta$ is shown in Fig.~\ref{fig:detpar}(a),
together with the relationship between $\MP$ and $\eta$ for $\wave_\theta=5$.
For $\eta=0.92$, Taylor vortices appear above $Re_{\rm TVF}\approx 146$.  For
these values of $\eta$, $\wave_z$, and $\MP$, Taylor-vortex flow remains
stable until $Re_{\rm WVF} \approx 201$, above which the flow becomes unstable
to wavy Taylor vortices.
Figure \ref{fig:vis1}(a) shows the Taylor-vortex flow
while Fig.~\ref{fig:vis2} shows the wavy-vortex flow, both computed at $Re=300$.

\section{Analysis in terms of Self-Sustaining Process}

We begin our analysis by introducing notation. Flow fields $\bU$ can be
decomposed as follows; see Fig.~\ref{fig:detpar}(b).
\begin{subequations}
\begin{eqnarray}
  \bU &=& \sum_k\sum_m \left(\hat{U}^{k,m}_r(r) \be_r+ \hat{U}^{k,m}_\theta(r) \be_\theta + \hat{U}^{k,m}_z(r) \be_z\right) e^{i(kz/\wave_z+m\MP\theta)}\qquad
  \label{eq:fourdecomp}\\
  &=& \bU_{\rm Cou}+\bU_{\rm mean} + \bU_{\rm roll} + \bU_{\rm streak} + \bU_{\rm wave}
\end{eqnarray}
\label{eq:decomp1}
\end{subequations}
where
\begin{subequations}
\begin{align}
\bU_{\rm Cou} &\equiv \left(A r + \frac{B}{r}\right)\be_\theta \label{eq:Cou}\\
\bU_{\rm mean} &\equiv \hat{U}^{0,0}_\theta(r) \be_\theta-\bU_{\rm Cou}\label{eq:mean}\\
\bU_{\rm roll} &\equiv \sum_{k\neq 0}
\left(\hat{U}_r^{k,0}(r)\be_r + \hat{U}_z^{k,0}(r)\be_z\right)\:e^{ikz/\wave_z}\\
  \bU_{\rm streak} &\equiv \sum_{k\neq 0}
  \hat{U}_\theta^{k,0}(r)\be_\theta \:e^{ikz/\wave_z}\\
  \bU_{\rm wave} &\equiv
\sum_k\sum_{m\neq 0}
%  \left(\hat{U}_r^{k,m}\be_r +\hat{U}_\theta^{k,m}\be_\theta + \hat{U}_z^{k,m}\be_z\right)\nonumber\\
  \hat{\bU}^{km}(r)
  e^{i(kz/\wave_z+m\MP\theta)}
  \label{eq:Ewave}
\end{align}
\label{eq:decomp2}
\end{subequations}
\noindent Note that \eqref{eq:mean} defines $\bU_{\rm mean}$ to be the 
$(\theta,z)$-independent deviation from laminar Couette flow $\bU_{\rm Cou}$,
in contrast to Waleffe
\cite{waleffe1995hydrodynamic,hamilton1995regeneration,waleffe1997self}
whose mode ${\rm M}$ includes the laminar Couette solution \eqref{eq:Cou}.
In terms of this decomposition, Taylor-vortex flow and wavy-vortex flow take the form
\begin{subequations}
\begin{align}
  \bU_{\rm TVF}&=\bU_{\rm Cou}+\bU_{\rm mean}+\bU_{\rm roll}+\bU_{\rm
    streak} \label{eq:TVF} 
  \\
  \bU_{\rm WVF}&=\bU_{\rm Cou}+\bU_{\rm mean}+\bU_{\rm roll}+\bU_{\rm
    streak}+\bU_{\rm wave}  
\end{align}
\label{eq:decomp3}
\end{subequations}
\noindent Waleffe's Self-Sustaining Process
\cite{waleffe1995hydrodynamic,hamilton1995regeneration,waleffe1997self}
describes three steps involving the components 
$\bU_{\rm roll}$, $\bU_{\rm streak}$, and $\bU_{\rm wave}$:\\
A) $\bU_{\rm roll} \Longrightarrow \bU_{\rm streak}$. This is a statement of
kinematic advection of the azimuthal velocity. \\
B) $\bU_{\rm streak} \Longrightarrow \bU_{\rm wave}$. This is described by
Waleffe as a linear instability. \\
C) $\bU_{\rm wave}\Longrightarrow \bU_{\rm roll}$.  The nonlinear interaction
of the wave with itself reinforces the rolls.

\subsection{Rolls to streaks}

\begin{figure}[t]
\centerline{
\includegraphics[width=0.75\columnwidth]{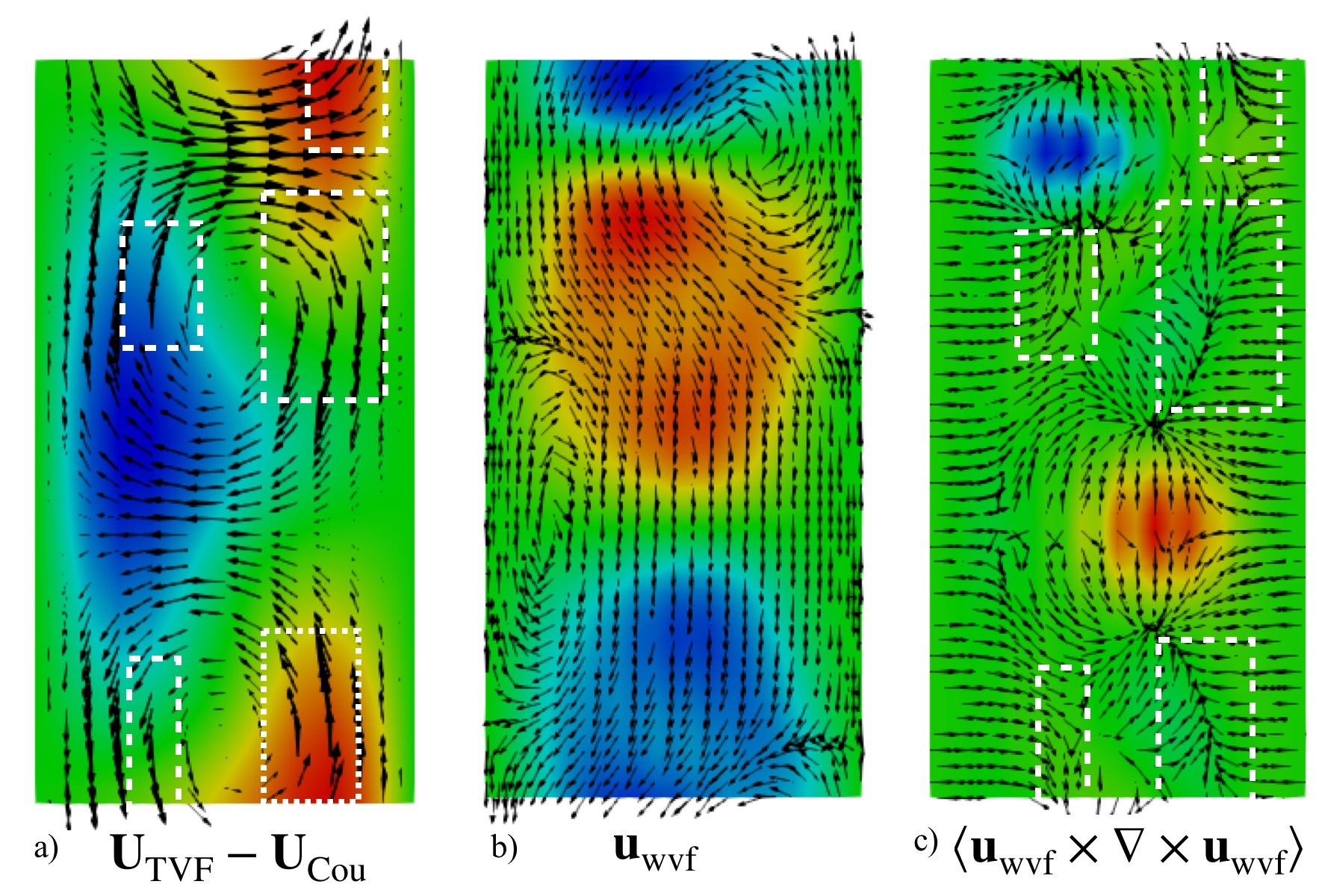}}
\caption{Visualizations in the meridional $(r,z)$ plane of (a) Taylor-vortex
  flow (without laminar Couette flow), (b) the $\MP=15$ eigenvector leading to
  wavy-vortex flow, and (c) nonlinear interaction of this eigenvector with
  itself. The parameters are $Re=300$ and $\eta=0.92$.
  The inner cylinder is on the left and the outer cylinder on the right.
  In each case, the
  meridional velocity within the plane is indicated by arrows and the
  azimuthal velocity perpendicular to it is indicated by colors.
  Red indicates a positive deviation of the azimuthal velocity
  from laminar Couette flow, blue a negative deviation, and green no deviation.
  Thus, in (a)
  the arrows show the rolls and the colors show the streaks of Taylor-vortex
  flow.  The white dashed boxes in (c) and (a) highlight the alignment 
  between the axial components (arrows) of $\langle\bu_{\rm
    wvf}\times\nabla\times\bu_{\rm wvf}\rangle$ and of the rolls of $\bU_{\rm TVF}$,
  which comprise the third step of the SSP.}
\label{fig:vis1}
\end{figure}

The SSP begins with streamwise invariant rolls $\bU_{\rm roll}$ and considers
the development of streaks from these rolls.
Rolls transport fluid with high azimuthal velocity
from the inner cylinder towards the outer cylinder and vice versa, 
causing the azimuthal velocity profile to vary along $z$
with the axial periodicity of the rolls.
In plane Couette flow, or
Waleffe's free-slip version \cite{waleffe1997self} now sometimes called
Waleffe flow \cite{beaume2014exact,beaume2015reduced,beaume2016modulated,
chantry2016turbulent,chantry2017universal}, rolls are not themselves an
equilibrium state. Hence in the planar case it is necessary to initiate the SSP
by inserting rolls into the flow and observing the resulting streak development.
Permanent rolls and streaks have been produced in variants of plane Couette flow
by including a spanwise-oriented wire or ribbon experimentally
\cite{dauchot1995finite,bottin1997intermittency,bottin1998experimental}
or numerically \cite{barkley1999stability}.
For the Taylor-Couette problem, however, this phase is straightforward.
The rolls and the streaks that they generate are contained in Taylor-vortex flow,
which bifurcates supercritically and exists as a
stable nonlinear equilibrium.  In Fig.~\ref{fig:vis1}(a), calculated at $Re=300$,
the rolls are the meridional-plane flow indicated by arrows.
The streaks are the axial variation in
the azimuthal flow driven by the rolls and are seen as the colored patches.

\subsection{Streaks to waves}

\begin{figure}[t]
\includegraphics[width=0.75\columnwidth]{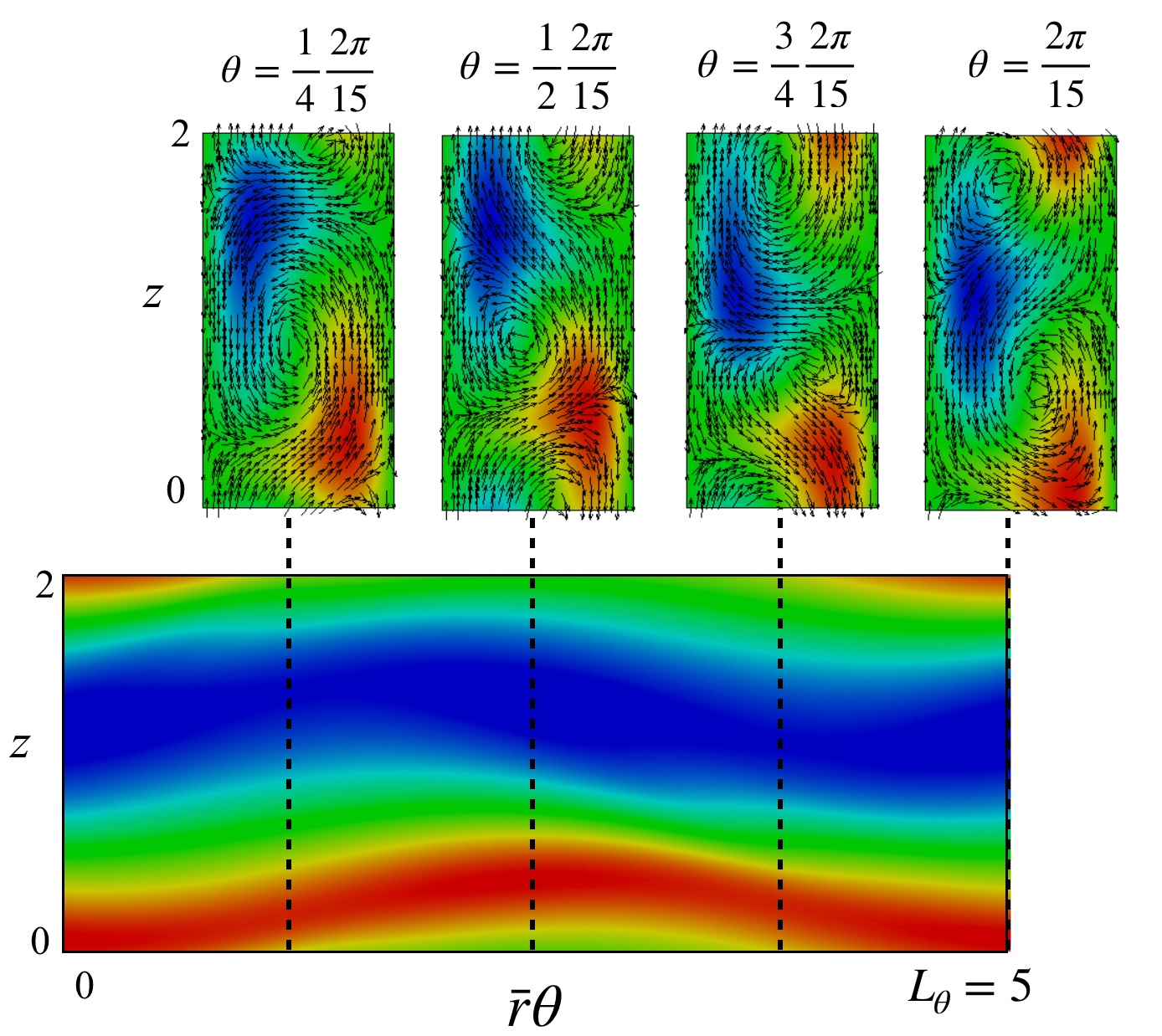}
\caption{Wavy-vortex flow (including Taylor-vortex flow but not
  laminar Couette flow) at $Re=300$.
  Above: four meridional planes over azimuthal period $[0,2\pi/15]$.
  Azimuthal velocity indicated by colors, meridional velocity by arrows.
  Below: one azimuthal period $[0,2\pi/15]$ at mid-gap $\bar{r}=12$.
  Radial velocity indicated by colors. The dashed lines indicate
  the positions of the four meridional planes shown above.
}
\label{fig:vis2}
\end{figure}

\begin{figure}[t]
\includegraphics[width=0.85\columnwidth]{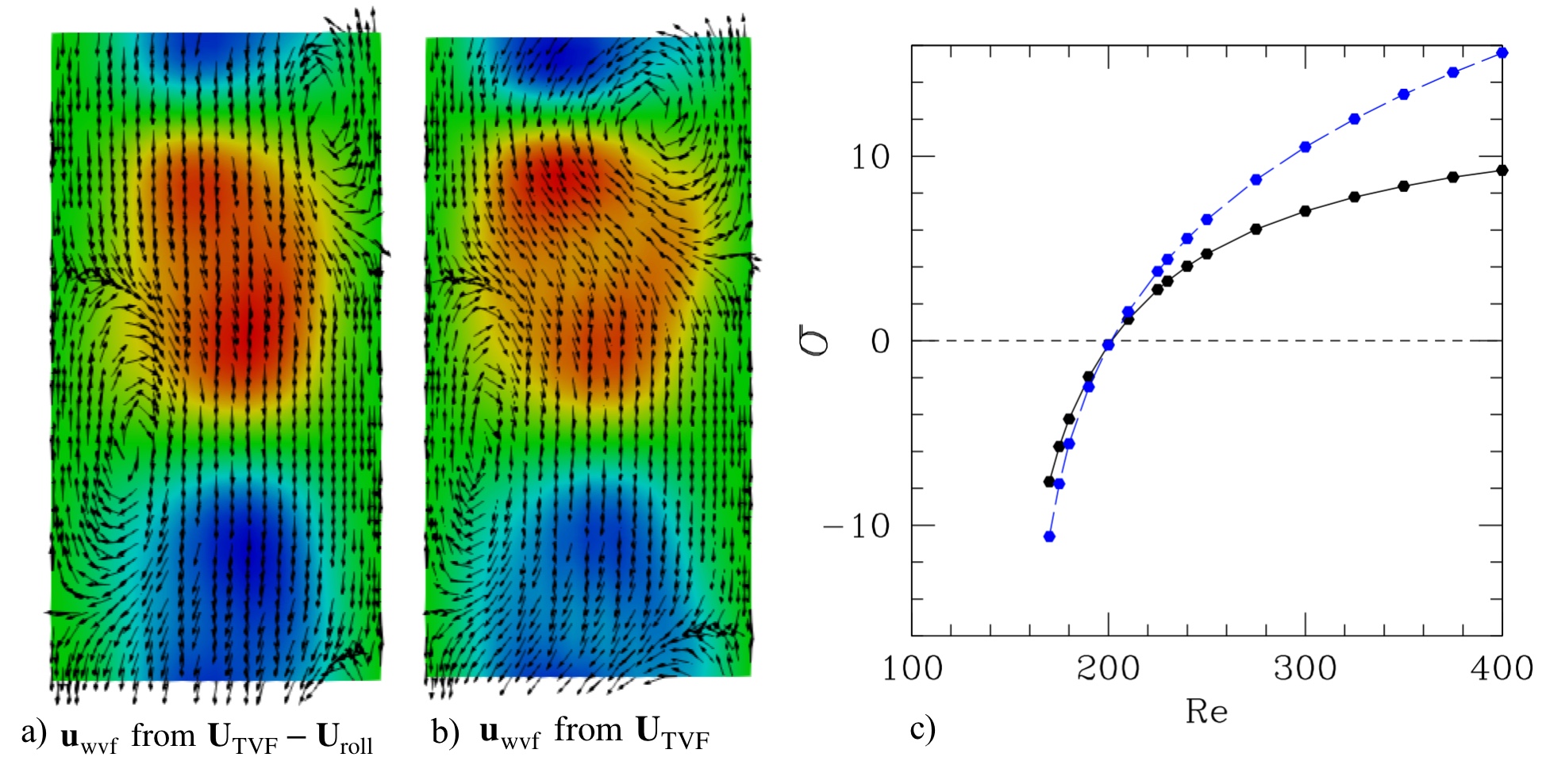}
\caption{Comparison of linearization about $\bU_{\text{TVF}}$ and
  about $\bU_{\text{TVF}}-\bU_{\text{roll}}$. Eigenvectors $\bu_{\text{wvf}}$
  resulting  from linearization about (a) only
  $\bU_{\text{TVF}}-\bU_{\text{roll}}$ and (b) the full
  $\bU_{\text{TVF}}$ for $Re=300$.  Azimuthal velocity designated by
  color (red for positive, blue for negative, green for zero) and
  radial and axial velocity by arrows. (c) Growth rate (real part of
  eigenvalue) for linearization about $\bU_{\text{TVF}}$ (black,
  solid), $\bU_{\text{TVF}}-\bU_{\text{roll}}$ (blue, dashed), as a
  function of $Re$. Since omitting $\bU_{\rm roll}$ from the
  base flow barely changes the eigenvector or eigenvalue, it is clear
  that it plays no role in the instability.}
\label{fig:eigcompare}
\end{figure}

We now turn to the second stage of the SSP in which the streaks become
unstable to waviness.
Once again, the situation in the Tayor-Couette problem is much more clear cut 
than in the planar case. The onset of waviness is a distinct supercritical
instability -- the transition from Taylor-vortex flow $\bU_{\rm TVF}$ to
wavy-vortex flow $\bU_{\rm WVF}$.
In the $\bU_{\rm WVF}$ state shown in Fig.~\ref{fig:vis2}, the flow
has azimuthal variation (waviness) and is an azimuthally travelling wave. 
In 1985, Jones \cite{jones1985transition} suggested that the instability arose
from the streaks, i.e.~the axial variation of the azimuthal flow, which he
called azimuthal jets.  Thirty years later, Martinand, Serre and
Lueptow~\cite{martinand2014mechanisms} confirmed this idea by constructing
the linear operator governing the wavy instability and showing that the
eigenvalues of the portion of the operator arising from the azimuthal shear,
i.e. the streaks, best matched the eigenvalues of the entire operator.  They
also demonstrated a number of common features between the transition to wavy
vortex flow and the Kelvin-Helmholtz instability, notably a phase speed
intermediate between that of the two cylinders and the multiplicity of
possible azimuthal wavenumbers.

We show this by a different procedure, carrying out linearization about
$\bU_{\rm TVF}$ and about $\bU_{\rm TVF}-\bU_{\rm roll}$, i.e. the Taylor
vortex flow without its radial or axial components; see
Eqs. \eqref{eq:decomp3}. Fig.~\ref{fig:eigcompare}
compares the eigenvectors and growth rates resulting from
these two linearizations. Since omitting $\bU_{\rm roll}$ from the base flow
barely changes the eigenvector or eigenvalue, it is clear that it plays no
role in the instability.  In contrast, linearization about $\bU_{\rm
  TVF}-\bU_{\rm streak}$, i.e. omitting the axial dependence of the azimuthal
flow, leads to eigenvalues with very small growth rate and eigenvectors with
no resemblance to those of $\bU_{\rm TVF}$. (These results are not displayed.)
These numerical experiments confirm that the instability mechanism responsible
for the transition of $\bU_{\rm TVF}$ to $\bU_{\rm WVF}$ is the axial
variation of the azimuthal velocity.

\begin{figure}[t]
\includegraphics[width=0.85\columnwidth]{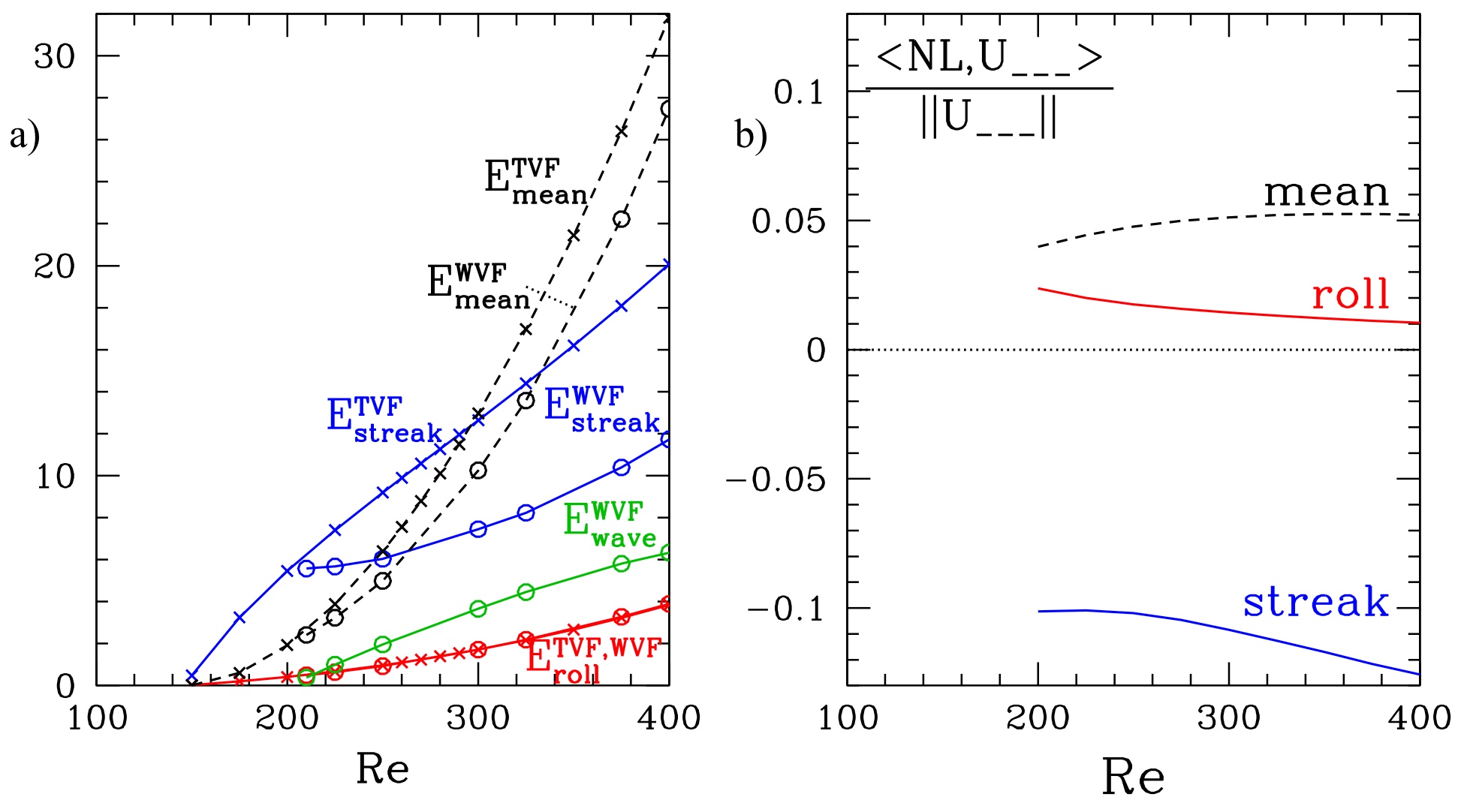}
\caption{(a) Energy decomposition for $\bU_{\rm TVF}$ and $\bU_{\rm WVF}$.
  (See Fig.~\ref{fig:detpar}(b) for definitions of this decomposition.)  Curves
  marked with crosses correspond to the components of $\bU_{\rm TVF}$
  originating at $Re=146$.  Curves marked with circles correspond to the
  energy components of $\bU_{\rm WVF}$ which bifurcates at $Re=201$.
%
%  The streak energy is lower for WVF than it is for TVF; the difference
%  between the two is almost precisely the energy in the waves (which is
%  necessarily zero for TVF).  
  The streak energy is lower for WVF than it is for TVF; the difference
  between the two is close to the energy in the waves (which is
  necessarily zero for TVF).  The energy in the deviation of the mean
  from Couette flow is also lower for WVF than for TVF.
  The energy in the rolls is approximately the same for the two flows.
%
%  (b) Inner product of nonlinear self-interaction of ${\bu}_{\rm wvf}$ with
%  rolls and with streaks.  The nonlinear term ${\bf NL} $ feeds the rolls and
%  drains the streaks.}
  (b) Normalized inner product of nonlinear self-interaction 
$\langle {\bf NL},{\bU}_{\rm roll}\rangle/||{\bU}_{\rm roll}||$, 
$\langle {\bf NL},{\bU}_{\rm streak}\rangle/||{\bU}_{\rm streak}||$ and
$\langle {\bf NL},{\bU}_{\rm mean}\rangle/||{\bU}_{\rm mean}||$ 
  for rolls, streaks, and deviation of the mean from Couette flow.
  The nonlinear term ${\bf NL}$ feeds the rolls and mean but
  drains the streaks.}
\label{fig:bifdiag}
\end{figure}

In addition to linearization, we examine the energy content in the flow
components of the nonlinear states.  We decompose both Taylor-vortex flow
$\bU_{\rm TVF}$ and the wavy-vortex flow $\bU_{\rm WVF}$ into components
given in Eqs.~\eqref{eq:decomp1}-\eqref{eq:decomp3} and compute the energy of each.
Fig.~\ref{fig:bifdiag}(a) shows the variation of the energy components as a
function of Reynolds number.
%(The energies of the Couette and mean components,
%$\bU_{\rm Cou}$ and $\bU_{\rm mean}$, are not shown.)
(The much larger energy of $\bU_{\rm Cou}$ and a contribution combining
${\bf U}_{\rm Cou}$ and ${\bf U}_{\rm mean}$ are not shown.)
$\bU_{\rm TVF}$ appears at $Re=146$ and $\bU_{\rm WVF}$ appears at $Re=201$.  
It can be seen that ${\rm E}^{\rm WVF}_{\rm streak}$, the energy of the
streaks in ${\bU}_{\rm WVF}$, is substantially decreased from the analogous quantity ${\rm
  E}^{\rm TVF}_{\rm streak}$ in $\bU_{\rm TVF}$. This decrease is almost exactly
counterbalanced by the energy in the waviness, ${\rm E}^{\rm WVF}_{\rm wave}$,
suggesting that the energy in the waviness is extracted from the streaks.
The energy in the rolls is small and is almost the same in the two states. 
Thus, in addition to the linear instability mechanism, the comparison between the 
energy content of the saturated nonlinear states with and without waves shows that streaks
feed the waves.
As stated by Waleffe \cite{waleffe1997self}, it is not the rolls but the streaks whose energy
is drained by the waves.

\subsection{Waves to rolls}

The key novelty of the SSP is the positive feedback of the waviness on the
rolls.  To study this in Taylor-Couette flow, we calculate the eigenvector
$\bu_{\rm wvf}$ responsible for the bifurcation to wavy vortices, shown in
Fig.~\ref{fig:vis1}(b).  (This complex eigenvector is shown here at one
spatial or temporal phase.)  We then compute the nonlinear interaction of
$\bu_{\rm wvf}$ with itself, in the form $\bu_{\rm wvf} \times \nabla \times
\bu_{\rm wvf}$.  Since $\bu_{\rm wvf} \sim e^{\pm i \MP \theta}$, this
quadratic term leads to azimuthal dependence of the form $e^{\pm 2 i \MP
  \theta}$ (second harmonic) and $1$ (constant).
We are interested in the constant contribution which has the form:
\begin{equation}
  {\bf NL} \equiv \langle \bu_{\rm wvf} \times \nabla \times \bu_{\rm wvf}\rangle
  \equiv \bu_{\rm wvf}^R \times \nabla \times \bu_{\rm wvf}^R
  + \bu_{\rm wvf}^I \times \nabla \times \bu_{\rm wvf}^I
  \label{eq:quad}\end{equation}
This term feeds back on the $\theta$-independent contributions $\bU_{\rm
  roll}$, $\bU_{\rm streak}$ and $\bU_{\rm mean}$.
A visualization of this vector quantity is shown in Fig.~\ref{fig:vis1}(c).
On a qualitative level, by comparing the arrows of Fig.~\ref{fig:vis1}(c) with
those of Fig.~\ref{fig:vis1}(a), 
one can see the feedback of this term on $\bU_{\rm roll}$.
The white-dashed boxes highlight regions in which the axial
component of the Taylor-vortex flow is strong and aligned with the axial component
of ${\bf \rm NL}$.
% Highlighted regions emphasize the resemblance between the meridional
%flow of NL and the rolls of Taylor-vortex flow.
The resemblance is especially strong on near-axial curves
in ${\bf NL}$ converging towards saddles above and
below regions with high azimuthal component shown in red.

A more quantitative picture of the feedback is presented in
Fig.~\ref{fig:bifdiag}(b). Shown is the normalized inner product between ${\rm
  NL} $ and each of
%$\bU_{\rm roll}$ and $\bU_{\rm streak}$.
$\bU_{\rm roll}$, $\bU_{\rm streak}$ and $\bU_{\rm mean}$ defined by:
\begin{align}
\langle {\bf NL},{\bf U}_{---}\rangle &=  \int_0^{L_z} dz \int_{r_1}^{r_2} r \:dr\: {\bf NL}(r,z)\cdot{\bf U}_{---}(r,z)
  \end{align}
where ${\bf U}_{---}$ is any of $\bU_{\rm roll}$, $\bU_{\rm streak}$ and $\bU_{\rm mean}$.
It can be seen
that ${\bf NL}$ has a positive overlap with ${\bf U}_{\rm roll}$, meaning that
indeed, the nonlinear interaction of $\bu_{\rm wvf}$ with itself acts as a
driving mechanism for rolls.
%(${\bf NL}$ also drives $\bU_{\rm mean}$, although this overlap is not shown on
%the figure.)
${\bf NL}$ also drives $\bU_{\rm mean}$.
In contrast, ${\bf NL}$ has a negative overlap
with ${\bf U}_{\rm streak}$ and hence this term tends to suppress the streaks.

\section{Conclusion}

According to the self-sustaining process (SSP) of \cite{waleffe1997self}, the
building block of transition to turbulence in plane Couette flow and other
wall-bounded shear flows, rolls induce streaks, which in turn undergo an
instability to waviness, whose nonlinear interaction feeds the rolls.  In
plane Couette flow, laminar flow (the analogue of ${\bf U}_{\rm Cou}$) is
stable for all Reynolds numbers; there is no equivalent of the steady
Taylor-vortex flow.  For Taylor-vortex flow, however, most of the steps of the
SSP are already in place. Vortices (rolls) induce streaks (axially periodic
variation of the azimuthal flow) kinematically via advection, as in plane
Couette flow.  We have confirmed that the instability to wavy-vortex flow is
due to this variation \cite{martinand2014mechanisms}.
In addition, we have shown that the energy of the waves in nonlinear wavy-vortex
flow compensates almost exactly for the decreased energy in the streaks, as compared to
the energy in the streaks of nonlinear Taylor-vortex flow.
The third step is the
feedback of the waves on the rolls, which is crucial for the SSP since in
plane Couette flow the rolls do not arise from a linear instability leading
to a nonlinear equilibrium.  We have shown that this feedback mechanism exists
in Taylor-Couette flow and that it is the rolls that are fed and not the streaks.
The nonlinear self-interaction of the waves generates localized regions with strong axial forcing:
this is the nature of the feedback on the Taylor vortices which closes the SSP.

\begin{acknowledgments}
This research was partly supported by the grant TRANSFLOW, provided by the Agence Nationale de la Recherche (ANR).
\end{acknowledgments}

\end{document}